     \documentstyle[12pt,epsf]{article}
     \newlength{\dinwidth}
     \newlength{\dinmargin}
\def\Journal#1#2#3#4{{#1} {\bf #2}, #3 (#4)}

\def\NPB{{\em Nucl. Phys.} B}
\def\PLB{{\em Phys. Lett.}  B}
\def\PRL{\em Phys. Rev. Lett.}
\def\PRD{{\em Phys. Rev.} D}
\def\ZPC{{\em Z. Phys.} C}
\def\lsim{\mathrel{\rlap{\lower4pt\hbox{\hskip1pt$\sim$}}
    \raise1pt\hbox{$<$}}}                
\def\gsim{\mathrel{\rlap{\lower4pt\hbox{\hskip1pt$\sim$}}
    \raise1pt\hbox{$>$}}}                
\begin{document}
\vspace*{10mm}
\begin{center}  \begin{Large} \begin{bf}
Study of polarized gluon distributions in diffractive reactions
at HERA\\
  \end{bf}  \end{Large}
  \vspace*{5mm}
  \begin{large}
S.V.Goloskokov
  \end{large}
\end{center}
Bogoliubov Laboratory of
Theoretical
  Physics, Joint Institute for  Nuclear Research, Dubna
  141980, Moscow
  region, Russia.
\begin{quotation}
\noindent
{\bf Abstract:} We consider the dependencies of  spin
asymmetries in the diffractive $J/\Psi$ and $Q \bar Q$
leptoproduction at HERA energies on the  structure of the
pomeron-proton coupling. It is shown that it is difficult to
study the spin structure of the pomeron coupling with the proton
from the $A_{ll}$ asymmetry. The $A_{lT}$ asymmetry is an
appropriate object for this investigation.
\end{quotation}

Now the pomeron nature is a problem of topical interest
due to the progress in analysis of diffractive processes at HERA
\cite{h1_zeus}. Investigation of the diffractive  vector meson
and $Q \bar Q$ production should give fertile information  on
the pomeron structure and on  the gluon distribution at small
$x$. The diffractive $J/\Psi$ and heavy $Q \bar Q$ production
have a significant role here. The predominant contribution in
such processes is determined by a color singlet $t$ -channel
exchange (pomeron). In the QCD-inspired models, the pomeron is
modeled by two gluons \cite{low}. The two--gluon couplings with
the proton in diffractive processes at small $x$ can be
expressed in terms of skewed gluon distribution in the nucleon
${\cal F}_X(X+\Delta)$ with $X\sim (Q^2+M_V^2)/W^2, \Delta \ll
X$, where $X+\Delta$ is a fraction of the proton momentum
carried by the outgoing gluon, and the difference between the
gluon momenta (skewedness) is equal to $X$ \cite{rad}.
Generally, these distributions are not connected with
ordinary gluon distributions.

The proposed HERA spin program gives a unique
possibility to study polarized  skewed gluon distributions in
the nucleon which are connected with the spin dependent pomeron
coupling. The spin structure of the pomeron  is an open
problem now.  In the QCD-based models, when the gluons
from the pomeron couple to a single quark in the hadron, the
effective pomeron coupling $V_{h I\hspace{-1.1mm}P}^{\mu}
=\beta_{h I\hspace{-1.1mm}P} \gamma^{\mu}$ appears which  looks
like a $C= +1$ isoscalar photon vertex \cite{lansh-m}.  The
spin-dependent pomeron coupling can be obtained if one
considers together with the Dirac form factor $\propto
\gamma^{\mu}$  the Pauli form factor \cite{nach} in the
electromagnetic nucleon current. We use in calculations the
following form of two--gluon coupling with the proton
\cite{golj}:
\begin{equation}
V_{pI\hspace{-1.1mm}P}^{\mu\nu}(p,t,x_P)= 4 p^{\mu} p^{\nu} A(t,x_P)
+(\gamma^{\mu} p^{\nu} +\gamma^{\nu} p^{\mu}) B(t,x_P).
\label{ver}
\end{equation}
Here $x_P$ is a fraction of
initial proton momenta carried by the pomeron. The term proportional to
$B$ represents the  pomeron coupling that leads to the
non-flip amplitude. The $A$ function  is the spin--dependent
part of the pomeron coupling that produces  spin--flip effects
nonvanishing at high-energies. The absolute value of the ratio
of $A$ to $B$ is proportional to the ratio of helicity-flip and
non-flip amplitudes. It has been found in \cite{gol_mod,gol_kr}
that  $\alpha=|A|/|B| \sim 0.1 -0.2 \,\mbox{GeV}^{-1}$ and has
a weak energy dependence.

A convenient tool to study the spin-dependent pomeron structure
might be polarized diffractive leptoproduction reactions. We
shall consider here the spin asymmetry of the $J/\Psi$ and $Q
\bar Q$ production. The cross section of these reactions has the
following important parts: leptonic and hadronic tensors and the
amplitude of the $\gamma^\star I\hspace{-1.7mm}P \to J/\Psi (Q
\bar Q)$ transition. The form of the leptonic tensor is quite
simple. The structure of the hadronic tensor for the vertex
(\ref{ver}) is discussed in Ref. \cite{golj}. The spin-average
and spin dependent cross sections with parallel and antiparallel
longitudinal polarization of a lepton and a proton are
determined by the relation $\sigma(\pm) = \left(
\sigma(^{\rightarrow} _{\Leftarrow}) \pm \sigma(^{\rightarrow}
_{\Rightarrow})\right)/2.$ These cross sections can be expressed
in terms of spin-average and spin dependent value of the lepton
and hadron tensors. The cross section of the $J/\Psi$
leptoproduction can be written in the form
\begin{equation}
\frac{d\sigma^{\pm}}{dQ^2 dy dt}=\frac{|T^{\pm}|^2}{32 (2\pi)^3
 Q^2 s^2 y}. \label{ds}
\end{equation}
For the spin-average  amplitude square we find \cite{golj}
\begin{equation}
 |T^{+}|^2=  N ((2-2 y+y^2) m_J^2 + 2(1 -y) Q^2) s^2 |B|^2
[|1+2 m \alpha|^2+|\alpha|^2 |t|] I^2.
\label{t+}
\end{equation}
Here $N$ is a known normalization factor $\alpha=|A|/|B|$ and
$I$ is the integral over transverse momentum of the gluon. The
term proportional to $(2-2 y+y^2) m_J^2$ in (\ref{t+})
represents the contribution of a virtual photon  with transverse
polarization. The $2(1 -y) Q^2$ term describes the effect of
longitudinal photons.\\[.2cm]
\begin{minipage}{8,3cm}
\epsfxsize=7.5cm
\centerline{\epsfbox{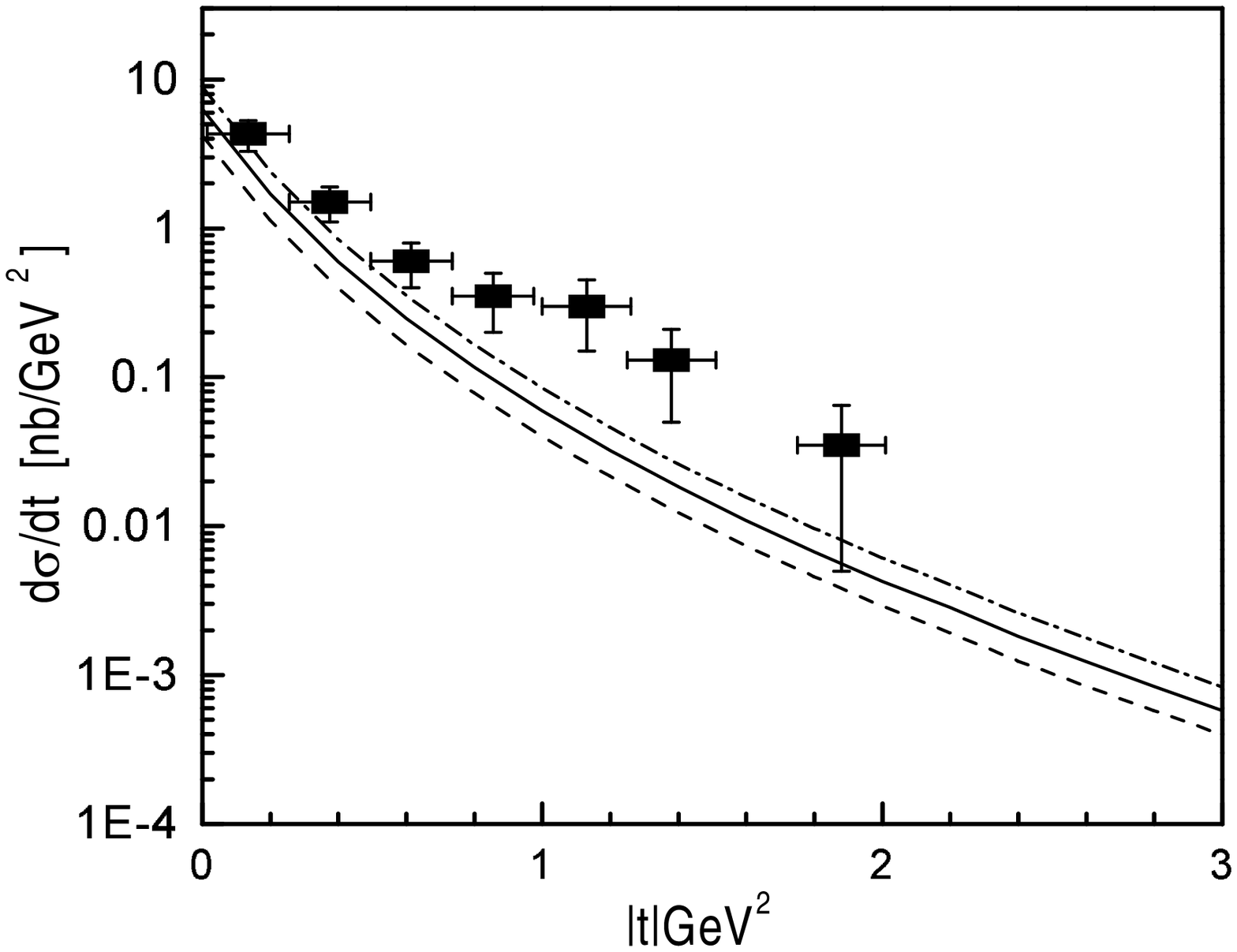}}
\end{minipage}
\begin{minipage}{0.24cm}
\phantom{aa}
\end{minipage}
\begin{minipage}{8.3cm}
\epsfxsize=7.5cm
\centerline{\epsfbox{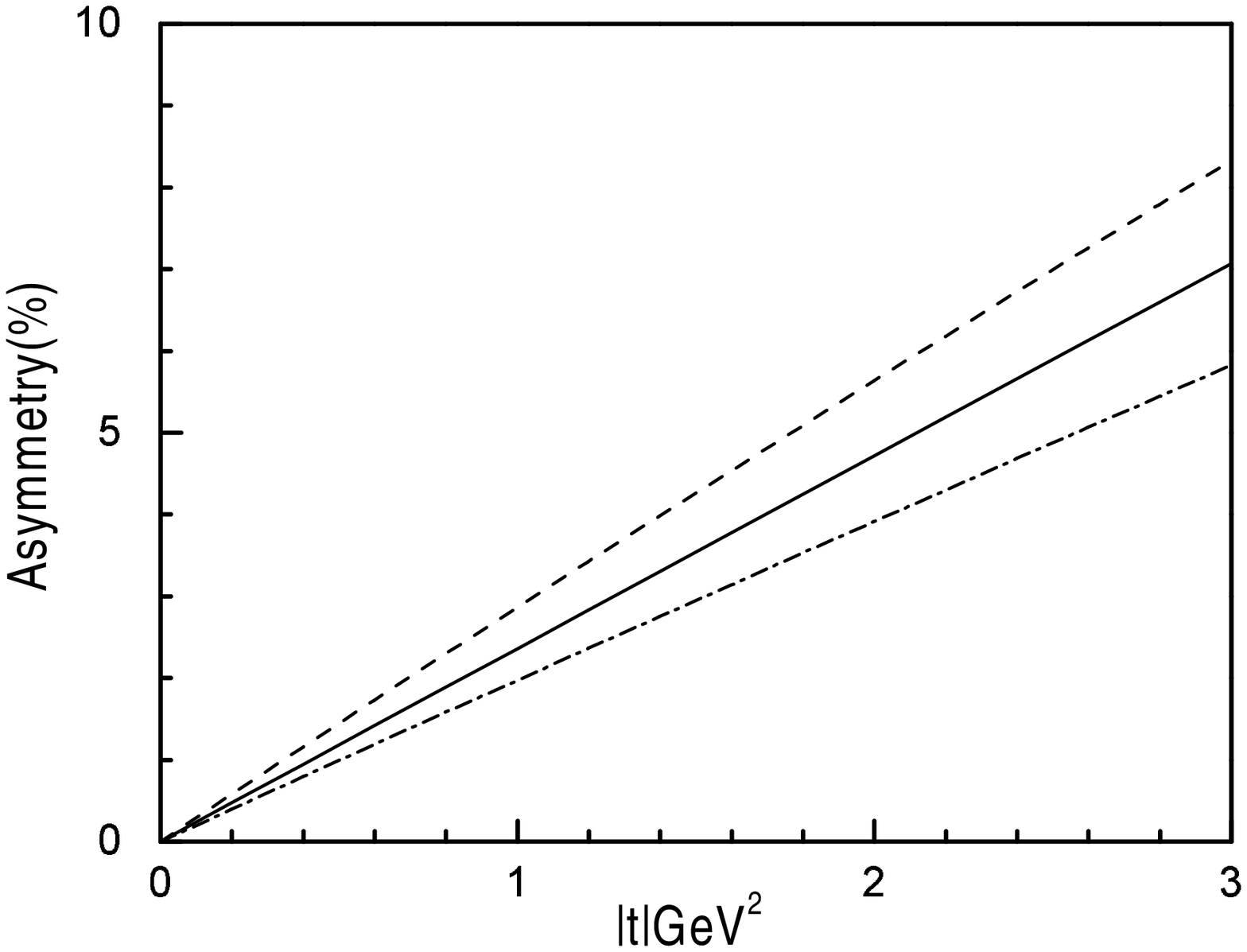}}
\end{minipage}
\medskip
\begin{minipage}{8cm}
Figure.1~ {\it The differential cross section of the $J/\Psi$ production
at HERA energy: solid line -for $\alpha=0$; dot-dashed line -for
$\alpha=0.1\mbox{GeV}^{-1}$;
dashed line -for $\alpha=-0.1\mbox{GeV}^{-1}$}.
\end{minipage}
\begin{minipage}{.95cm}
\phantom{aaa}
\end{minipage}
\begin{minipage}{8cm}
Figure.2~ {\it The predicted $A_{ll}$ asymmetry of the $J/\Psi$ production
at HERMES: solid line -for $\alpha=0$; dot-dashed line -for
$\alpha=0.1\mbox{GeV}^{-1}$;
dashed line -for $\alpha=-0.1\mbox{GeV}^{-1}$}.
\end{minipage}

We shall integrate the cross sections (\ref{ds}) over $y$ and
$Q^2$ with $Q^2_{max} \sim 4 \mbox{GeV}^2$
\begin{equation}
\frac{d\sigma^{\pm}}{dt}=\int_{y_{min}}^{y_{max}} dy
\int_{Q^2_{min}}^{Q^2_{max}} dQ^2
\frac{d\sigma^{\pm}}{dQ^2 dy dt}.
\end{equation}
The cross section for the $J/\Psi$ production at HERA energy
$\sqrt{s}=300GeV^2$
is shown in Fig.1.  The spin-average cross sections
are sensitive to $\alpha$ but the shape of all curves is the
same. Thus, it is difficult to extract information about the
spin--dependent part of the pomeron coupling from the
spin--average cross section of the diffractive vector meson
production.

Similar calculation has been done for the spin--dependent part
of the cross section. As a result, the following form of
asymmetry is found \cite{golj}:
\begin{equation}
\label{all_a}
A_{ll}= \frac{\sigma(-)}{\sigma(+)} \sim \frac{|t|}{s}\frac{(2-\bar y)
(1+2 m \alpha)}
{(2-2\bar y+\bar y^2)\left[ (1+2 m \alpha)^2+\alpha^2|t| \right]
}.
\end{equation}
The predicted asymmetry of the $J/\Psi$ vector meson production
for  HERMES as a function $\alpha$ is shown in Fig.\ 2. The
asymmetry is equal to zero in the forward direction.  Thus, the
ordinary gluon distribution $\Delta G$ cannot be extracted from
$A_{ll}$ in agreement with the results of \cite{mank}. The
predicted asymmetry  does not vanish for nonzero $|t|$.  The
value of the asymmetry for $\alpha \neq 0$ is dependent on  the
$A$--term of the pomeron coupling. However, the sensitivity of
the asymmetry to $\alpha$ is not very strong. Thus, it will not
be so easy to study the spin structure of the pomeron coupling
with the proton from the $A_{ll}$ asymmetry of the diffractive
$J/\Psi$ production.

Similar calculations have been done for the heavy $Q \bar Q$
diffractive leptoproduction. As previously, we have included in
our analyses the graphs where the gluons from the pomeron couple
to a different quark in the loop as well to the single one.
This provides the gauge-invariant scattering amplitude. The
obtained $A_{ll}$ asymmetry is proportional to $x_p$ like for
the vector meson production and has a weak energy dependence
 (here $x_p$ is typically of about $.05-.1$). The
cross section integrated over the pomeron momentum transfer was
calculated because the recoil proton is usually not detected in
diffractive experiments. The obtained asymmetry is quite small
and does not exceed 1-1.5\%. It has a week $\alpha$ dependence
 and does not vanish for $\alpha=0$ as in the $J/\Psi$
production. The estimated $Q^2$ dependence of $A_{ll}$ asymmetry
of the diffractive open charm ($c \bar c$) production is shown
in Fig. 3. Note that the smallness of the asymmetry is caused by
the strong cancellation in $\Delta \sigma$ between  the graphs
where the gluons couple to the single and a different quark in the
loop. Similar calculation for the Donnachie-Landshoff model of
the pomeron, where only the planar graphs appear, provides the
asymmetry which is about 10\%. \cite{gol_all}\\[.1cm]
\begin{minipage}{8,3cm}
\epsfxsize=8.2cm
\centerline{\epsfbox{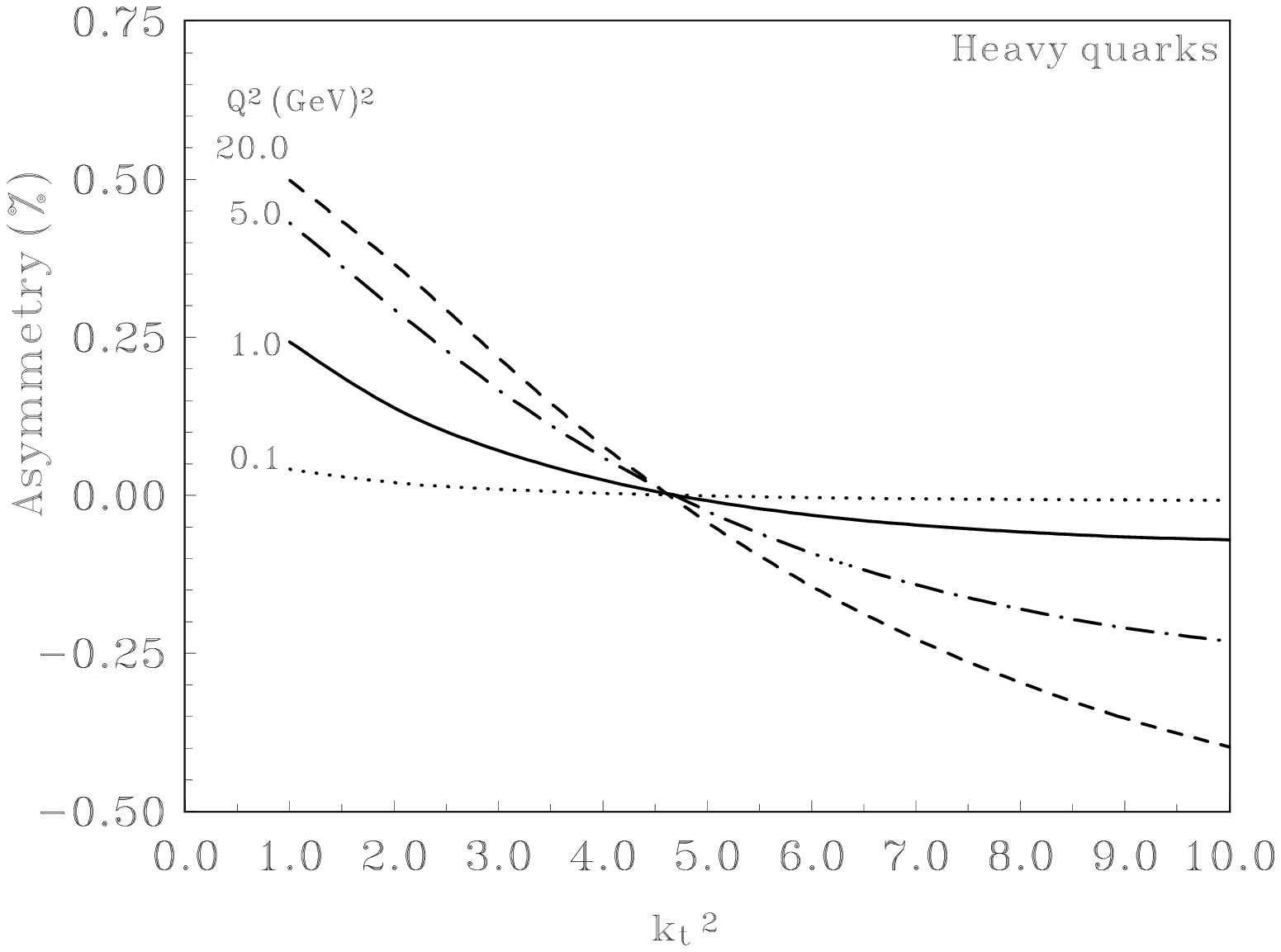}}
\end{minipage}
\begin{minipage}{0.24cm}
\phantom{aa}
\end{minipage}
\begin{minipage}{8.2cm}
\epsfxsize=8.3cm
\centerline{\epsfbox{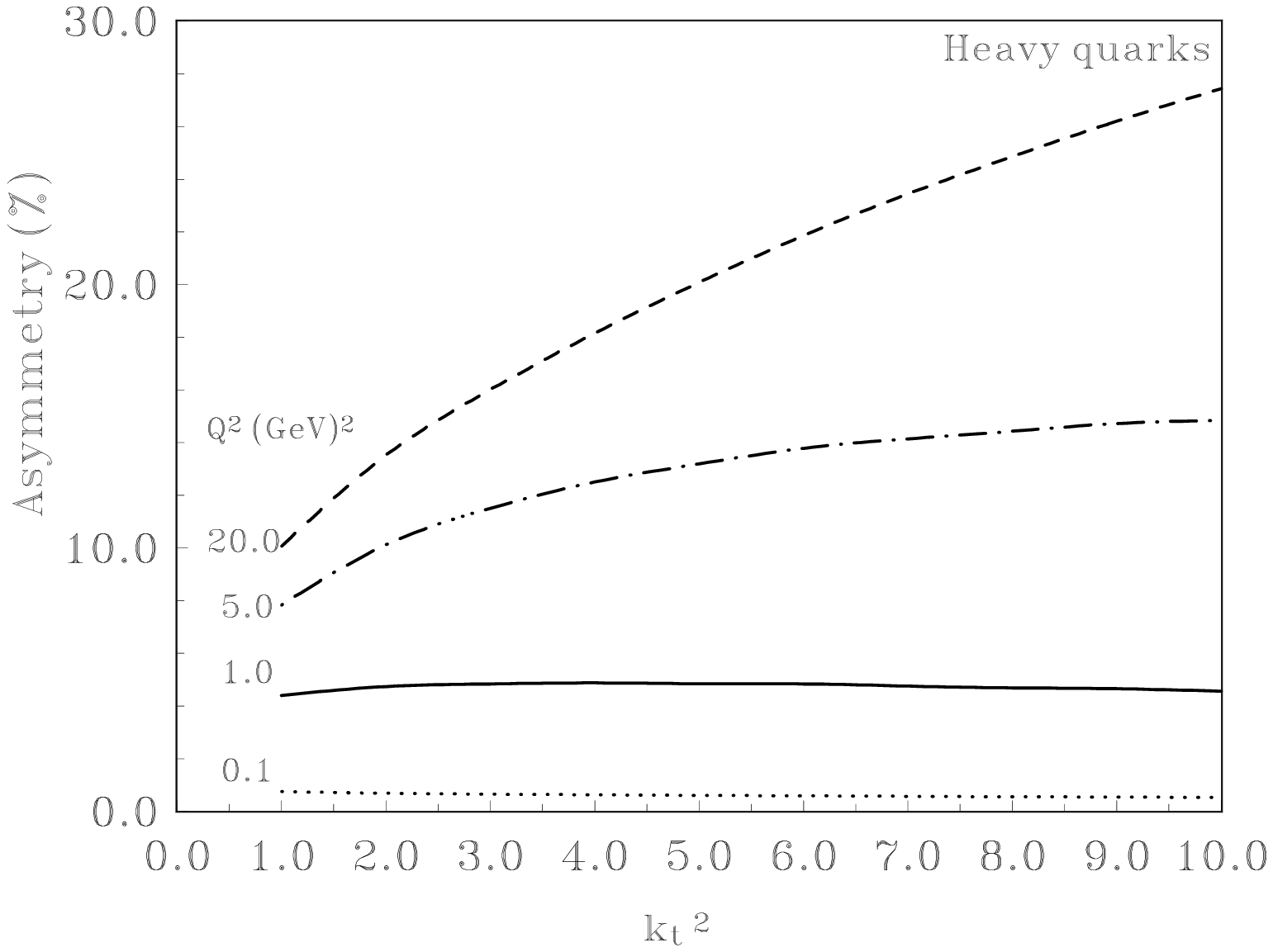}}
\end{minipage}
\bigskip
\begin{minipage}{8cm}
Figure.3~ {\it The predicted $Q^2$ dependence of $A_{ll}$ asymmetry
for the $c \bar c$ production at HERA for $\alpha=0.1 \mbox{GeV}^{-1}$, $x_p$=0.1, $y$=0.5}.
\end{minipage}
\begin{minipage}{.95cm}
\phantom{aaa}
\end{minipage}
\begin{minipage}{8cm}
Figure.4~ {\it The predicted $Q^2$ dependence of $A_{lT}$ asymmetry
for the $c \bar c$ production at HERA  for $\alpha=0.1\mbox{GeV}^{-1}$, $x_p$=0.1, $y$=0.5}.
\end{minipage}

Another important object which can be studied at polarized
HERA is the $A_{lT}$ asymmetry with longitudinal lepton and
transverse proton polarization. It has been found that $A_{lT}$
asymmetry is proportional to the scalar production of the proton
spin vector and the jet momentum. Thus, the asymmetry integrated
over the azimuthal jet angle is zero. We have calculated the
$A_{lT}$ asymmetry for the case when the proton spin vector is
perpendicular to the lepton scattering plane and the jet
momentum is parallel to this spin vector.  This is impossible
explicitly in the experiment  and the integration over some
region of the angle between spin vector and jet momentum should
be done. Then, the additional factor
$\int_{-\Theta}^{\Theta} \cos(\theta) d\theta/
\int_{-\Theta}^{\Theta}  d\theta$ appears which is about 0.98
for $\Theta=20^{o}$. The estimated $Q^2$ dependence of the
$A_{lT}$ asymmetry integrated over $t$ for
$\alpha=0.1\mbox{GeV}^{-1}$ is shown in Fig. 4. The predicted
asymmetry is huge and has a strong $k^2_{\perp}$ dependence. The
large value of $A_{lT}$ asymmetry is caused by the fact that it
does not have a small factor $x_p$ as a coefficient.

In the present report, the polarized cross section of the
diffractive hadron leptoproduction at high energies has been
studied.  As a result, connection of the spin--dependent cross
section in the diffractive production with the pomeron coupling
has been found. Generally, the function $B$ should be determined
by the spin--average and the function $A$ - by the polarized
skewed gluon distribution in the proton. We predict not
small value of the $A_{ll}$ asymmetry of the diffractive vector
meson production at the HERMES energy. The predicted $A_{ll}$
asymmetry in the $Q \bar Q$ leptoproduction is smaller than 1.5\%.
The nonzero asymmetry for $\alpha=A/B=0$  is completely
determined by the $\gamma^{\alpha}$ term in the pomeron coupling
(\ref{ver}). The $A_{ll}$ asymmetry in diffractive processes for
nonzero momentum transfer has been found to be dependent on the
$A$ term of the pomeron coupling which has a spin dependent
nature. However, the sensitivity of  asymmetry to the
$\alpha$--ratio  is quite weak. Thus, the $A_{ll}$ asymmetry in
diffractive reactions is not a good tool to study the polarized gluon
distributions of the proton and spin structure of the pomeron.
Otherwise, it has been found not small $A_{lT}$ asymmetry in
diffractive $Q \bar Q$ production. This asymmetry is
proportional to $\alpha$ and can be used to obtain direct
information about the spin--dependent part of the pomeron
coupling $A$. The $A_{lT}$ asymmetry is an appropriate object to
investigate the spin structure of pomeron coupling and the
skewed polarized gluon structure of the proton.

We can conclude that the pomeron coupling structure
can be analysed in  diffractive processes.
So, the important test of the spin structure of QCD at
large distances can be carried out by studying diffractive
reactions in future polarized experiments  at HERA.

This work was supported in part by the Heisenberg-Landau Grant.
The author is grateful to the Organizing Committee for financial support.

\end{document}